\documentclass[preprint,12pt]{elsarticle}

\usepackage{graphicx}         
\usepackage{bm}               
\usepackage{amssymb}          
\usepackage{amsmath}          

\usepackage{color}            

\def \msun        {M_\odot}

\begin{document}
\newcommand{\aea}{Astron. Astrophys.}
\newcommand{\apj}{Astrophys. J.}
\newcommand{\apjl}{Astrophys. J. Lett.}
\newcommand{\cqg}{Class.  Quant. Grav.}
\newcommand{\grg}{Gen.  Rel. Grav.}
\newcommand{\jcap}{J. Cosmol. Astropart. Phys.}
\newcommand{\mnras}{Mon. Not. R. Astron. Soc.}
\newcommand{\prd}{Phys. Rev. D}
\newcommand{\prl}{Phys. Rev. Lett}
\newcommand{\plb}{Phys. Lett. B}

\begin{frontmatter}

\title{Dynamical Evidence of a Dark Solitonic Core of $10^{9}M_\odot$ in the Milky Way.}

\author[1]{Ivan De Martino\footnote{E-mail: ivan.demartino@dipc.org} }
\author[1,2,3]{Tom Broadhurst\footnote{E-mail:tom.j.broadhurst@gmail.com}}
\author[4]{S.-H. Henry Tye\footnote{E-mail: iastye@ust.hk} }
\author[5,6]{Tzihong Chiueh\footnote{E-mail: chiuehth@phys.ntu.edu.tw}}
\author[7]{Hsi-Yu Schive\footnote{E-mail: hyschive@gmail.com}}

\address[1]{Donostia International Physics Center (DIPC), 20018 Donostia-San Sebastian (Gipuzkoa)	Spain.}
\address[2]{Department of Theoretical Physics, University of the Basque Country UPV/EHU,
             E-48080 Bilbao, Spain}
\address[3]{Ikerbasque, Basque Foundation for Science, E-48011 Bilbao, Spain}
\address[4]{Institute for Advanced Study and Department of Physics, Hong Kong University of Science and Technology, Hong Kong}
\address[5]{Department of Physics, National Taiwan University, Taipei 10617, Taiwan}
\address[6]{National Center for Theoretical Sciences, National Taiwan University, Taipei 10617, Taiwan}  
\address[7]{National Center for Supercomputing Applications, Urbana, IL 61801, USA}


\begin{abstract}

A wavelike solution for the non-relativistic universal dark matter (wave-DM) is rapidly gaining interest, following pioneering simulations of cosmic structure as an interference pattern of coherently oscillating bosons. A prominent solitonic standing wave is predicted at the center of every galaxy, representing the ground state solution of the coupled Schr\"odinger-Poisson equations, and it has been identified with the wide, kpc scale dark cores of common dwarf-spheroidal galaxies. 
A denser soliton is predicted for Milky Way sized galaxies where momentum is higher, so the de Broglie scale of the soliton is smaller, $\simeq 100$ pc, of mass $\simeq 10^9 M_\odot$. Here we show the central motion of bulge stars in the Milky Way implies the presence of such a dark core, where the velocity dispersion rises inversely with radius to a maximum of $\simeq 130$ km/s, corresponding to an excess central mass of $\simeq 1.5\times 10^9 M_\odot$ within $\simeq 100$ pc, favouring a boson mass of $\simeq 10^{-22}$ eV. This quantitative agreement with such a unique and distinctive prediction is therefore strong evidence for a light bosonic solution to the long standing Dark Matter puzzle. 

\end{abstract}


\end{frontmatter}

\section{Introduction}

The nature of the Dark Matter (DM) is understood to require new physics, as baryonic matter described by standard particle physics is found to contribute only 17\% of the cosmic mass density \cite{Planck16,Cyburt2016}. We know DM is non-relativistic, to the earliest limits of observation, otherwise the Cosmic Microwave Background (CMB) and the large scale distribution of galaxies would be featureless on small scales. In addition, collisions between galaxy clusters show no detectable self-interaction other than gravity \cite{Marikevitch2004, Clowe2006}. However, the composition of this collisionless, ``Cold Dark Matter" (CDM) is unclear, particularly with the continued laboratory absence of any new heavy particles to stringent limits \cite{lux2017}. 

Furthermore, predictions of CDM are problematic on small scales, $<10$ kpc, in relation to low mass galaxies. An interesting alternative to CDM is dark matter composed of an extremely light boson ($m_\psi \sim 10^{-22}$ eV), sometimes called Fuzzy Dark Matter (FDM) \cite{Widrow1993,Sin1994,Goodman2000,Hu2000,2006PhLB..642..192A}. Since de Broglie wavelength of these light bosons is of order $\lambda_{dB} \sim 1$ kpc, the physics differs fundamentally from CDM on scales below $\lambda_{dB}$. The uncertainty principle means bosons cannot be confined within this de Broglie scale, naturally suppressing dwarf galaxy formation below $10^{10} M_\odot$. So the question arises, what replaces the cuspy inner profile of CDM within 
10 kpc? It turns out that there is a rich, unforeseen granular sub-structure on the de-Broglie scale, as revealed by the simulations \cite{Schive2014}, with qualitative \cite{Mocz2017} and quantitative support \cite{Veltmaat2018} by subsequent independent simulations.
Wave-DM cosmological  simulations of \cite{Schive2014, Schive2014b} that evolve classical scalar fields obeying the non-relativistic Schr\"odinger-Poisson equations  \cite{Widrow1993,Sin1994,Goodman2000}, under the simplest assumption of negligible self-interaction other than gravity, produce halos
with a central core that is a stationary, minimum-energy solution, sometimes called a soliton, surrounded by a granular envelope resembling a CDM halo when averaged azimuthally \cite{Schive2014,Schive2014b}. With the boson mass as the only free parameter, the soliton solution has a scaling symmetry, thus forming a one parameter family of solutions. We shall refer to FDM with this one-parameter solitonic core as wave-DM or ``$\psi$DM".

Simulations with a wide dynamical range from the de Broglie wavelength to cosmological scale must rely on some simplifying modifications which suppress small scale cosmological perturbations, and do not result in the formation of axions miniclusters \cite{2019arXiv191207064N}. A comparison between the Schr\"odinger-Poisson equation and  the Gross-Pitaevskii equation points out the loss of phenomena due to the intrinsic quantum nature of the problem \cite{Sikivie2009,Erken2012}. As an example, the formation of caustic ring substructures in the galactic plane related to the existence of vorticities in the axion Bose-Einstein condensate \cite{Sikivie2009}, which may suppress the dark matter density near the galactic center \cite{Banik2013}. Recently, it has also been shown that for even for ultra light boson with masses $\sim10^{-22}$, quantum effects lead to the formation of those caustic rings \cite{Banik2017}, but these will be absent in the wave-DM  assumed here to  behave purely  classically,  so the formation of caustic rings which are related to the formation of quantum vortexes \cite{Sikivie2009} is suppressed. 

The most distinctive feature of wave-DM is the formation of one prominent solitonic wave at the base of every virialised potential \cite{Schive2014,Schive2014b,Schive2016}, representing the ground state solution, where self gravity of the condensate is matched by effective pressure from the uncertainty principle.  The solitons found in the simulations have flat cored density profiles that accurately match the known time independent solution of the Schr\"{o}dinger-Poisson equation
\cite{Schive2014,Schive2014b}, for which the soliton mass scales inversely with its radius \cite{Guzman2006}. A boson mass of $\sim 10^{-22}$eV has been derived in this context, corresponding to a solitonic core of 1 kpc of mass $10^8 M_\odot$, by matching the phase space distribution of stars within the well studied Fornax galaxy \cite{Schive2014}, that is representative of the most common class of dwarf spheroidal galaxies, where dark matter dominates over stellar mass. Furthermore, the relatively weak dynamical friction of wave-DM, compared to CDM, can help to account for the otherwise puzzling presence today of ancient globular clusters at large radius in this Fornax galaxy \cite{Hui2016}. A denser solitonic core is predicted for wave-DM in more massive galaxies, like the Milky Way, using a scaling relation derived from the simulations between the mass of soliton and its host virial mass, $m_{soliton} \simeq m_{virial}^{1/3}$ \cite{Schive2014,Schive2014b,Schive2016,Veltmaat2018} so that a soliton mass of $\simeq 10^9M_\odot$ has been predicted for the Milky Way, with a radius of 100 pc \cite{Schive2014}.  Here we explore the central dynamics of giant stars in the Milky Way that have increasing numbers of accurate velocities from which it has been possible to derive usefully accurate velocity dispersion profiles within a few degrees from the center of the Galaxy, as a function of both Galactic Latitude and Longitude.

{ Theoretical motivation for such light bosons is
clearly provided by the  generic axions in String Theory
\cite{Cicoli2012,Marsh2014,Tye20017}, with the likely possibility of a very wide ranging discrete mass spectrum for the dark matter, limited by black hole physics \cite{Stott2018}, which empirically we may conclude is dominated by the universal dark matter on a scale of $10^{-22}$ eV, with the possibility of other less significant contributions accounting for claimed dark cores of globular clusters \cite{Emami2018}.} 

\section{Bulge Stellar Dynamics and wave-DM}\label{sect:data}

The dynamics of stars and gas clouds in the in the Milky Way's bulge has been recently investigated 
in \cite{Zoccali2014,Schonrich2015,Portail2017}. Such studies have pointed out the complexity of the dynamics of the different components constituting the bulge. Red Clump (RC) stars have been used to map the bulge. Specifically, Zoccali et al. (2014) \cite{Zoccali2014} uses spectra of $\sim 5000$ RC stars, which were collected by the Giraffe Inner Bulge Survey at the ESO-VLT with the spectrograph FLAMES, to derive radial velocity, metallicities and dispersion velocities. Recently, Portail et al (2017) \cite{Portail2017} have investigated a combination of data from VVV, UKIDSS and 2MASS infrared surveys together with kinematics data from the BRAVA and OGLE surveys to reconstruct the velocity field in the Milky Way's bulge  in the whole bulge area $|l| < 10^\circ$ and $-10^\circ < b < +5^\circ$. Both Zoccali et al. (2014) \cite{Zoccali2014} and Portail et al (2017) \cite{Portail2017} have found in the data an enhancement of the dispersion velocity in the innermost region of the Milky Way's bulge. In Portail et al (2017), N-body simulations have been employed to investigate the causes of such a velocity excess, and used to demonstrated the need of the presence of an additional $10^9M_\odot$ point-like mass in the centre of the Milky Way in order to account for the dispersion velocity peak in the region $|l| < 5^\circ$ and $b < +5^\circ$. Nevertheless, there is no evidence of such a point-like mass arising from other observational studies of the galactic center, all claiming a central point-like mass of the order of $\sim 10^6 M_\odot$ \cite{Ghez2008, 2017ApJ...837...30G}.

To construct a model of the inner dispersion velocity, we start with the well understood components: the central supermassive black hole, the bulge, and the disk. The supermassive black hole is the least important component for our purposes, since it affects stars orbiting within a few parsecs where the gravitational potential, $\Phi_{BH}$, is usually assumed to be generated by a point-like mass $M_{BH}=(4.5\pm0.6)\times 10^6M_{\odot}$ \cite{Ghez2008}. The bulge dominates the inner region of the galaxy with an estimated scale radius, $r_b \sim 500$ pc and density profile that has been most recently carefully determined by \cite{Portail2017} to be well approximated as $\rho_b = \rho_{b,0} e^{-r/r_b}$, where $\rho_{b,0}$ is readily computed using the total mass of the bulge given in Table 2 of \cite{sofue2012} resulting in $ \rho_{b,0}=14.27 M_\odot$pc$^{-3}$. The galactic disk has little importance  on the sub-kpc scale of interest here, but we include for completeness. It is usually described with an exponential disk \cite{BT} with central surface mass density
$\Sigma_0=8.44\times10^2 \rm{M}_\odot \rm{pc}^{-2}$, 
and scale length $R_d = 3.5$ kpc \cite{sofue2009, sofue2013}
with a corresponding gravitational potential 
in the plane of the disk ($z=0$), $\Phi(R)\equiv\Phi(R, z=0)$.

 We now add to above visible components a dark matter contribution, following the $\psi$DM simulation results of  \cite{Schive2014, Schive2014b} which comprises a solitonic standing wave core in the central region, surrounded by an extended halo with $\mathcal{O}(1)$ density fluctuations on the de-Broglie scale, but is simply described radially by a NFW-like profile when azimuthally averaged \cite{Schive2014}.

The density profile of the solitic core is well approximated by:
\begin{equation}\label{eq:sol_density}
\rho_c(r) \sim \frac{1.9~(m_\psi/10^{-23}~{\rm eV})^{-2}(r_c/{\rm kpc})^{-4}}{[1+9.1\times10^{-2}(r/r_c)^2]^8}~M_\odot {\rm pc}^{-3}.
\end{equation}
where  $m_\psi$ is the boson mass and $r_c$ is the core radius of the solitonic solution given by the derived scaling of the boson mass with the total halo mass:
$\propto m_{\psi}^{-1}M_{halo}^{-1/3}$ derived by \cite{Schive2014,Schive2014b}. For this purpose we adopt a Milky Way $M_{halo}^{MW}=2\times10^{12}M\odot$, again following \cite{Portail2017} so the gravitational potential generated by DM halo density distribution is: 

\begin{equation}\label{eq:potbulge1}
 d\Phi_{DM}(r) = G \frac{M_{DM}(r)}{r^2} dr\,,
\end{equation}
using as boundary condition  $\Phi_{DM}(\infty)=0$, and computing the mass as
\begin{equation}\label{eq:massbulge1}
 M_{DM}(r) = 4\pi\int_0^r x^2 \rho_{DM}(x)dx\,.
\end{equation}
In the above equation, $ \rho_{DM}$ is the total dark matter density profile which, as stated above, is given by the solitonic core profile plus the very well-known NFW density profile describing the region out of the soliton.
Finally, the total gravitational potential becomes,
\begin{equation}
 \Phi(r) = \Phi_{BH}(r)+\Phi_{B}(r)+\Phi_{disk}(R)+\Phi_{DM}(r)\,.
\end{equation}
 In practice, 
 the only significant contributions 
 within the 500 pc region of interest here are the bulge and soliton.
 Therefore, we drop the disk potential in our calculations and compute the dispersion velocity  integrating the spherical Jeans equation:

\begin{equation}
 \frac{d(\rho_*(r)\sigma_r^2(r))}{dr} = -\rho_*(r)\frac{d\Phi(r)}{dr}-2\beta\frac{\rho_*(r)\sigma_r^2(r)}{r},
\end{equation}
where $\rho_*(r)$ is the tracer RC star density, and $\Phi(r)$ is the total gravitational potential 
generated by the enclosed mass. The observed RC stars are assumed to be proportional to the density profile  of this combination of bulge plus inner Plummer profile and are used as the tracer component for the Jeans equation as this assumption should be a good first approximation.

The anisotropy parameter $\beta$ characterizes the orbital structure of 
the halo, and it is 
(\cite{BT}, Equation (4.61))
zero for an isotropic system. This is a good approximation for the region of interest. 
The scale length of the bulge is much larger than the central 100 pc scale of interest here, so the bulge stars fall through the central region approximately equally from all directions. Therefore, for sake of simplicity, we can assume isotropy being also supported by measurements of bulge stars dynamics finding $\beta \sim 0$ out to $\sim 15$ kpc \cite{King2015, Kafle2012, Deason2013}. Finally, to directly compare the dispersion velocity with the data, we project the velocity dispersion along the line of sight as follows
\begin{equation}
\sigma^2_{los} (R) = \frac{\int_{R}^{\infty} \sigma^2_r(r)\rho_*(r)(r^2-R^2)^{-1/2} r dr}{\int_{R}^{\infty} \rho_*(r)(r^2-R^2)^{-1/2}rdr}\,.
\end{equation}

The above model may be regarded as the minimal model in the context of wave-DM and the results are described below and shown  in Figures 1\&2 providing a good fit to the data.  

The above model is clearly able to generate the excessive dispersion detected in the data, but with a hint of a slight overestimate by the model in the very center on a scale of $\simeq 100$ pc. This may be understood with a small and self consistent refinement to model (dashed curves in Figures 1\&2) arising from bulge stars that have become "captured" by the central soliton and hence see a relatively small interior mass thereby lowering the projected central velocity dispersion when averaged together with the majority of bulge stars. Without detailed numerical models, it is hard to predict the proportion of such soliton bound stars, so we simply estimate empirically the possible level of central moderation of the central velocity dispersion with the addition of a small stellar core \cite{Chen2017,Lin2018} by adding an additional Plummer distribution for such bound stars with a scale radius $r_p\sim 150$ pc to approximately match the mean soliton core considered in this work:
\begin{equation}
\rho_p =  \frac{\rho_{p,0}}{\left(1+\frac{r^2}{r_p^2}\right)^{5/2}}\,,
\end{equation}
where $\rho_{p,0}$ is computed so that the densities of the Plummer sphere and the bulge match at $r\sim 15$ kpc, resulting in $ \rho_{p,0}=0.05 M_\odot$pc$^{-3}$. The results as shown as dashed curves in Figures 1\&2, where as expected the presence of such small orbits within the flat core of the the soliton leads to a net reduction of the velocity dispersion at some level that increases in significance towards the center, as discussed further below.

In addition to the above we have also compared our simplified bulge model with the more complex N-body treatment by Portail et al \cite{Portail2017} who reproduce well the  detailed bulge triaxiality responsible for the X-shape orbits beyond $500$ pc in their simulations. For this purpose we include the same DM contribution as Portail et al. \cite{Portail2017}, which differs from the wave-DM contribution above by increasing in density DM all the way to the center instead of solitonic core. This pure NFW contribution adds on average about 10 km/s to the central velocity dispersion near the center, as shown in Figures 1\&2 and discussed further below.  

\section{Results}

We compute the dispersion velocity projected along the line of sight from a system composed by a central black hole, a bulge, and a $\psi$DM halo  
(with the DM profile given by the Eq. \eqref{eq:sol_density}). Then, we solve the Jeans equation for different values of the boson mass in the range $[6 - 14]\times 10^{-23}$ eV. Let us remark that we only vary the boson mass in our model while we kept fix the other parameters to their best fit values \cite{Portail2017}. This strategy is designed to clearly demonstrate that the central enhancement on the dispersion velocity is due to only the presence of the soliton.

\begin{figure}[!ht]
\centering
\includegraphics[width=1.0\columnwidth]{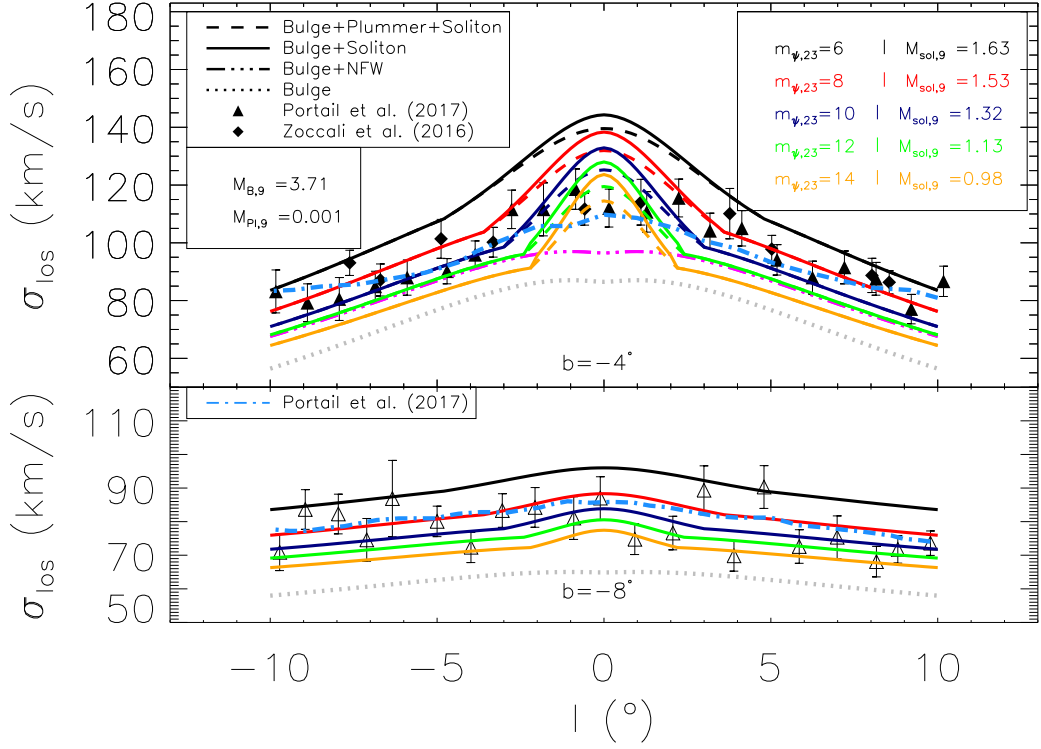} 
\vspace{-0.5cm}
\caption{Predicted dispersion velocity as function of Galactic longitude for the two sets of data of latitude of $b=-4^\circ$ (upper panel) and latitude $b=-8^\circ$ (lower panel). In both panels predictions are made for different values of the boson mass which correspond different values of the soliton mass. In the legend, $m_{\psi,23}$  means the boson mass in units of $10^{-23}$ eV, and $M_{sol,9}$  indicates the  soliton mass in units of $10^9 M_\odot$. Solid lines represent our baseline model accounting for dark matter, bulge and the central black hole, while the dashed lines also account for a denser star population described by a Plummer density profile. The triple dot-dashed magenta line represents the prediction of a NFW DM halo plus bulge model, while the grey dotted line is just the bulge model, which both fall well below the measured central dispersion, demonstrating the need for unseen additional matter.  Finally, the dot-dashed light blue line is the best fit model from Portail et al. (2017) \cite{Portail2017} which includes all galactic component plus a significant additional unseen point-like mass of $\sim10^9 M\odot$ that Portail etal add to obtain the central enhancement seen in the data. In the figure we indicate the total mass of the Bulge and of the Plummer sphere in units of $10^9 M_\odot$, computed at $500$ pc and labeled as $M_{b,9}$ and $M_{pl,9}$ respectively.}\label{fig1}
\end{figure}

In the upper panel of Fig. \ref{fig1} we compare the dispersion velocity profile as function of longitude, as predicted from our model, 
with the data from Zoccali et al. 2014 \cite{Zoccali2014} at latitude of $-4^\circ$. 
Notice that, in our model, the Plummer component of bulge stars does influence the best fit by mildly reducing the expected central dispersion. Therefore, although this extra component were added to the Portail et al. model, it does not affect the need for significant additional unseen mass that we can explain with our solitonic core. In the lower panel of Fig. \ref{fig1}, we compare our predicted dispersion velocity profiles with a set of data at latitude of $b=-8^\circ$. At such latitude, our simple model converge to the more complete one from Portail et al. (2017) \cite{Portail2017} favouring the boson mass $m_\psi\sim10^{-22}$ eV.
 
\begin{figure}[!ht]
\centering
\includegraphics[width=0.99\columnwidth]{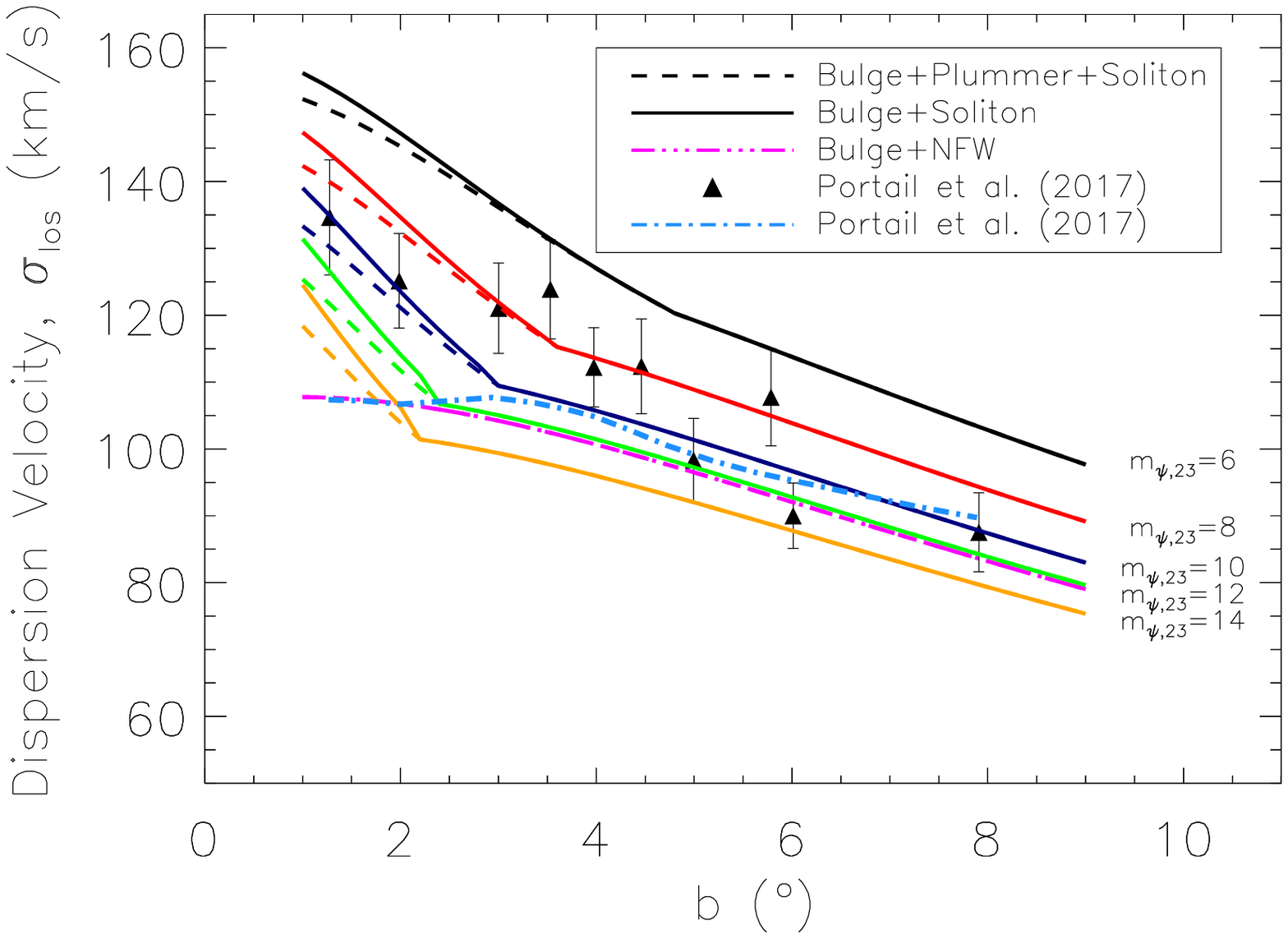}
\caption{Predicted dispersion velocity as function of Galactic latitude for a fixed longitude $l=0^\circ$. Again our baseline model (dotted magenta curve), with no soliton, underpredicts the central velocity dispersion. Our prediction is in good agreement with the independent prediction of Portail et al. (2017) shown as light blue that includes a detailed simulation of the stellar bulge, central black hole 
and a disk and NFW halo for the Galaxy, and who also conclude on this basis the need for a significant additional mass of $\simeq10^9M_\odot$. As in Fig. \ref{fig1}, the solid lines are our baseline model, and the dashed lines show the effect of including a Plummer density distribution for a small fraction of the
visible stars contributing he to central dispersion due to the presence of a soliton. population}\label{fig2}
\end{figure}

In Fig. \ref{fig2} we show the predicted dispersion velocity profile as function of Galactic latitude, and compare it 
with the data from Portail et al. 2017 \cite{Portail2017}. Lines and colors follow the same labelling of Fig. \ref{fig1}. Again our baseline model (triple dots-dashed magenta curve), with no soliton, underpredicts the central values of the velocity dispersion and is in good agreement with the independent prediction of Portail et al. (2017) \cite{Portail2017} (dash-dotted blue curve), who also conclude from this 
model shortfall that a significant additional central mass must be added.

For both of these independent data sets we see the
need for an additional central mass, which is unseen and in excess of the visible bulge by $1.5\times 10^9 M_\odot$. Thus, we carried out a $\chi^2$-test varying only the boson mass.  
Our analysis favours $m_{\psi,23}=9\pm0.6$ eV and $m_{\psi,23}=8.8\pm0.6$ eV for our baseline model without the additional Plummer density distribution of stars and with such additional density profile, respectively.
This conclusion is quantitatively in good agreement with the careful dynamical modelling by Portail et al. (2017) \cite{Portail2017}.  Clearly the combination of the gravitational contributions of bulge, disk  and NFW DM halo, are not enough to reproduce the observed dispersion velocity in the inner part of the Milky Way. The amount of additional matter differs a little from that derived from Portail et al. (2017) who reconcile the difference adding a compact mass of the order of 
$10^9 M_\odot$, whereas our soliton component is extended over $\simeq 100$ pc, depending on the associated boson mass, which acts as a point mass to a good approximation as the vast majority of
the tracer RC stars lie outside this 100 pc radius, following the observed bulge scale length of 500 pc.
Note that the known net rotation of the bulge has no bearing on our estimate of the velocity 
dispersion for the spherical Jeans equation. What may be significant is the possibly compression of  the soliton
by the surrounding potential of the bulge mass, which may lead to a higher bulge mass for a given 
boson mass, as more wave-DM is forced to relax into the ground state comprising the soliton by up to a factor of $\simeq 2$ in the exploration of this gravitational coupling by Chan et al. (2018) \cite{chan2018}. However, this possible enhancement in density leaves the product of the soliton mass and radius unchanged as
the inverse scaling of mass with radius is guaranteed by the soliton solution, but the corresponding
boson mass is raised somewhat above value that fits 
best in Figures 1 \& 2, as the soliton mass is enhanced and the radius compressed somewhat. 

\subsection{Comparison with other constraints and validity of the assumptions}

Let us devote this section to the discussion of possible shortcomings in our analysis, and the comparison with constraints deriving from other independent analyses.

Verification of the standard CDM interpretation of DM as a heavy non-relativistic particle has not emerged from stringent direct and indirect particle detection experiments due to its  weakly interacting nature \cite{lux2017,Xenon1T}. Astrophysical and cosmological observations are also clarifying the nature of DM, with the phase-space density of typical dwarf galaxies requiring that dark matter as a fermion must have a mass larger than 400 eV \cite{Tremaine1979}.
Bosons can be much lighter for example the QCD axion, or the much lighter axion-like solution favored here. 
Galaxy observations on small scales provide another means of defining the nature of DM, that at face value disfavour a heavy collisionless particles, given the common presence of central cores in dwarf galaxies, unless "feedback" from gas motion is somehow able to influence the dark matter \cite{Oh2011}. The lack of detection of  the number of low mass halos remains a major difficulty for CDM  \cite{Moore1999,Boylan-Kolchin2012}, but searches
are underway for small dark halos that may be revealed by lensing \cite{Hezaveh2016}.  These difficulties have led to the exploration of dark matter, that is either "warm" (WDM), meaning relativistic at early times, or self-interacting rather than purely collisionless (SIDM). The first proposition would generate a core in dwarf galaxies, while the latter would solve the lack of missing low mass halos \cite{Viel2005,Rocha2013,Bulbul2014}. However, it seems not possible solve both issues at the same time \cite{Maccio2012}.

As already mentioned in the previous sections, ultra-light bosons with mass of the order of $\sim10^{-22}$ eV are a very promising candidate to solve all the small scale issues of the standard CDM. Primary CMB and CMB-lensing fluctuations have been used to obtain a lower limit on the boson mass, $\geq 10^{-23}$ eV \cite{Hlozek2015, Hlozek2017}. Nevertheless, the study of the small scale structure have limited the boson mass to the range $ [0.4, 6]\times10^{-22}$ eV \cite{Schive2014,Calabrese2016,Chen2017,Marsh2015}.
This is supported by the discovery of unprecedentedly diffuse galaxies orbiting the Milky Way with  central DM densities that are far lower, and cores that are much wider, than the corresponding one in the context of the standard CDM \cite{Torrealba2016,Torrealba2019, Broadhurst2019}, where the opposite behaviour is expected for dwarf galaxies. 

The clustering properties of hot gas at high redshift that gives rise to the Lyman-$\alpha$ forest has been used to constrain DM models. On small scales it has been claimed that there is a marginal excess of power relative to the predictions of WDM for a particle mass $\simeq 4$ keV by the free streaming scale in this context.  Lyman-$\alpha$ bounds strongly depend on  thermal histories and the underlying reionisation model which may stultify
such constraints \cite{Garzilli2017, Hui2016}. Furthermore this comparison relies on idealised simulations that do not include the known complexities of common galaxy gas outflows at high redshift \cite{Frye2001, Pettini2001} and also small scale variation in ionizing sources from stars and AGN at high redshift which doubtless boost the variance in the measured Ly-$\alpha$ power spectrum.
The WDM based constraint  on the free streaming scale  \cite{Baur2016} have been quickly translated to a lower limit for the boson mass of $\geq 7\times10^{-21}$ eV relying on the analogy on the relatively sharpe power spectrum cut-off inherent to the Jeans scale in the Wave-DM context.  However, the rich small scale structure revealed by Wave-DM simulations \cite{Schive2014,Mocz2017,Veltmaat2018} may be expected to boost structure particularly in the high momentum infall regions of filaments surrounding early galaxies in this context. It has been pointed out more generally that differences in the matter power spectrum between ultra light boson and warm dark matter can play a fundamental role \cite{Hui2016}. Indeed, recent N-Body cosmological simulations of ultra-light bosons accounting for the hydrodynamics of the inter-galactic medium  have been compared with the SDSS Lyman-$\alpha$ spectrum leading to the conclusion that boson masses in the range ($0.1-2.1$]$\times10^{-21}$ eV are excluded. These results leave a window for masses below the $10^{-22}$ eV where quantum effects may have an impact on the cut-off of the matter power spectrum \cite{Irsic2017a, Irsic2017b,Armengaud2017, Nori2019}. 
Therefore, our estimation of the boson relying on the kinematics of stars in the Milky Way's Bulge is not in tension with the most recent analysis of Lyman-$\alpha$ spectrum in the context of ultra-light bosons. Moreover, it worth to note this claimed tension can also be readily alleviated in recent axion wave-like cosmological simulation using a more general axion potential that deviates mildly from simple harmonic oscillation, demonstrating that quantum effects may readily reduce such tension allowing for  boson with mass $\sim10^{-22}$ eV \cite{Schive2019}.

A lesson learned from the wave-DM cosmological  simulations is that solitons form first followed by the relatively slow virialization of the outer halo \cite{Schive2014b,Schive2016}, and with a clear relationships between the soliton mass and halo mass $M_{sol} \propto M_{halo}^{1/3}$. Despite an inherent scatter in this relation the slope does not  seem to evolve significantly, so that despite the merging history of halos this relation is maintained. More investigations of the effect of non baryonic matter is required to understand how the soliton and soliton-halo relation may be affected by the accumulation of stars near the center of galaxies - which may well be significantly determined by the presence of the soliton. We have speculated that the formation of bulges and of super massive black holes may be better understood in the wave-DM context. What we can say at this point is that there is little visible matter within the central 100 pc of interest here - as the stellar bulge has a scale length of 2 kpc, and the gas and stars visible within 100 pc amount to several $10^7 M_\odot$, only a small fraction of the $10^9 M_\odot$ we determine here for the core. Hence we find it unlikely that the central soliton is significantly compressed by this additional matter but, beyond the scope of current work,  full cosmological simulations including both gas and stellar contributions are required in the future \cite{Schive2014,Schive2016}.

\section{Discussions and Conclusions}

 The deep spectroscopic measurements of the extincted central bulge stars have provided a surprisingly large velocity dispersion rising to $130$ km/s, well above the baseline bulge contribution $\simeq 80$ km/s. We have shown that this excess can be accounted for by the
a central dark mass predicted by Schive et al. (2014)  in their pioneering $\psi$DM simulations of wave-DM \cite{Schive2014, Schive2014b}, that reveal a rich interference pattern on the de Broglie scale with a prominent standing, soliton wave expected at the center of every halo. 
The scale of this core for the Milky Way has been predicted by Schive et al. (2014)  \cite{Schive2014}, to be  $\sim 100$ pc, with a mass of $\simeq 10^9M_\odot$, based firmly on the derived soliton scale for the large core of the Fornax dwarf spheroidal galaxy, which provided the associated boson mass for this model. 

This is a vulnerable prediction unique to wave-DM that requires a new massive and large central mass component in the Milky Way. Nevertheless it would appear from our analysis that such a dark object is viably present at the center of the Galaxy, helping to explain the excessive central velocity dispersion of bulge stars observed. The effect of this central mass manifests itself very simply in the data as an essentially Keplerian decline in the velocity central few 100 pc region because the bulge stars
that feel this central excess mass  mostly lay at radii exceeding this 100 pc radius as a $1/r$ decline, in good agreement with the data. Furthermore, this result is in good agreement with the
careful numerical modelling of the central bulge star dynamics by Portail et al. (2017) \cite{Portail2017}, who construct a fully dynamical model of the bulge, finding a need for a similar additional, large compact mass of $\simeq 2\times  10^9 M_\odot$, in good agreement with our conclusions, both in terms of the amount of mass and the need for a relatively compact object.

 Independently, measured large central velocities of $\simeq 100$ km/s on a scale of $\simeq 100$ pc have also been made on the basis of orbiting streams of gas around the center of the Milky Way \cite{Molinari2011,Kruijssen2015,Henshaw2016} with a tentative light disk of stars with linearly rising rotation reaching $\simeq 100$ km/s at a limit of 100 pc, \cite{Schonrich2015}. 
There is also the real possibility that a central Milky Way soliton may be directly detected from Compton oscillations of the bosonic scalar field \cite{Khmeinitsky:2013lxt} that induce a General Relativistic time modulation of clocks with a cyclical modulation amplitude of $\simeq 200$ ns on a timescale of months \cite{idm2017,idm2018}, and may be detected by SKA which has the sensitivity to identify suitable millisecond pulsars within the central soliton radius of the Milky Way, where the oscillation amplitude is strongest.  Hence, we can look forward to increasingly illuminating data for testing this unique soliton signature of wave-DM, with the capability to provide definitive evidence for a light bosonic solution to the long standing Dark Matter puzzle.


\begin{thebibliography}{99}
 
 \bibitem{Planck16}
 Planck Collaboration, \aea\, {594}, A13 (2016)
 
 \bibitem{Cyburt2016}
R.H. Cyburt  et al., Rev. Mod. Phys. 88, 015004 (2016)

\bibitem{Marikevitch2004}
M. Markevitch, A.H. Gonzalez, D. Clowe, A. Vikhlinin, W.Forman, C. Jones, S. Murray, W. Tucker,
\apj, 606, 819-824 (2004)

\bibitem{Clowe2006}
D. Clowe, M. Bradac, A.H. Gonzalez, M. Markevitch, S.W. Randall, C. Jones, D. Zaritsky, \apj, 648, L109-L113 (2006)

\bibitem{lux2017} 
LUX Collaboration, Phys. Rev. Lett. 118, 021303 (2017) 

\bibitem{Widrow1993}
L. M. Widrow, N. Kaiser, \apj, 416, 71 (1993)

\bibitem{Sin1994}
Sang-Jin Sin Phys. Rev. D 50, 3650 (1994)

\bibitem{Goodman2000}
J. Goodman, New Astronomy Reviews, 5, 103 (2000)

\bibitem{Hu2000}
W. Hu, R. Barkana, A. Gruzinov, \prl, 85, 1158-1161 (2000)

\bibitem{2006PhLB..642..192A} 
L. Amendola,  R. Barbieri, (2006), Physics Letters B, 642, 192

 \bibitem{Schive2014}
Hsi-Yu Schive, T. Chiueh, T. Broadhurst, 
Nature Physics , 10,  7, 496-499 (2014)

\bibitem{Mocz2017}
P. Mocz et al., \mnras, 471, 4559-4570 (2017)

\bibitem{Veltmaat2018}
J. Veltmaat, J.C.  Niemeyer, B. Schwabe, (2018), arxiv:1804.09647

\bibitem{Schive2014b}
Hsi-Yu Schive, M-H Liao, T-P Woo, S-K Wong, T. Chiueh, T. Broadhurst, and W-Y. Pauchy Hwang
\prl \, 113, 261302 (2014)


\bibitem{2019arXiv191207064N} 
Niemeyer, J.~C.\ 2019, arXiv e-prints, arXiv:1912.07064


\bibitem{Sikivie2009}
P. Sikivie and Q. Yang,
Phys. Rev. Lett. 103,  111301, (2009)

\bibitem{Erken2012}
O. Erken,  P. Sikivie, H. Tam and Q. Yang, Phys. Rev. D 85,063520, (2012) 

\bibitem{Banik2013}
N. Banik, P. Sikivie  Phys. Rev. D 88, 123517 (2013)

\bibitem{Banik2017}
N. Banik et al., Phys. Rev. D 95, 043542  (2017) 





\bibitem{Schive2016}
Hsi-Yu Schive, T. Chiueh, T. Broadhurst, K.-W. Huang, 
\apj, 818, 89 (2016) 

\bibitem{Guzman2006}
F.S. Guzman, L.A. Ure\~{n}a-L\'{o}pez,
\apj, 645, 814-819 (2006).

\bibitem{Hui2016}
L. Hui,  J. P. Ostriker, S. Tremaine, E. Witten,
\prd, 95, 043541 (2017)

\bibitem{Cicoli2012}
M. Cicoli, M.D. Goodsell, A. Ringwald,
Journal of High Energy Physics, 2012, 146 (2012)

\bibitem{Marsh2014} 
D.J.E. Marsh, J. Silk, 
\mnras, 437, 2652-2663 (2014)

\bibitem{Tye20017}
S.-H. H. Tye, S. S.C. Wong, 
Journal of High Energy Physics, 2017, 94 (2017)

\bibitem{Stott2018}
M.J. Stott,  J.E. Marsh, David, (2018), arXiv:1805.02016

\bibitem{Emami2018}
R. Emami, T. Broadhurst, G. Smoot, T. Chiueh, L. Hoang Nhan, (2018) arXiv:1806.04518

\bibitem{Zoccali2014}
M. Zoccali, O.A. Gonzalez, S. Vasquez, V. Hill, M. Rejkuba, E. Valenti, A. Renzini, A. Rojas-Arriagada, 
I. Martinez-Valpuesta, C. Babusiaux, T. Brown, D. Minniti, A. McWilliam, 
\aea\,, 562, A66  (2014) 

\bibitem{Schonrich2015}
R. Sch\"{o}nrich, M. Aumer, S.E. Sale, \apj\,, 812, 2, L21 (2015)

\bibitem{Portail2017}
M. Portail, O. Gerhard, C. Wegg, M. Ness, 
\mnras\,, 465, 2, 1621-1644 (2017).

\bibitem{Ghez2008} A. M. Ghez, S. Salim, N. N. Weinberg, J. R. Lu, T. Do, J. K. Dunn, K. Matthews, M. Morris, S. Yelda, E. E. Becklin, T. Kremenek, M. Milosavljevic, J. Naiman, \apj\,, {689} (2), 1044 (2008)

\bibitem{2017ApJ...837...30G} S. Gillessen , et al., \apj, 837, 30 (2017)



 
 \bibitem{sofue2012}
Y. Sofue, PASJ, 64, 2 (2012)
 
\bibitem{BT}
J. Binney, \& Tremaine, S. 2008, Galactic Dynamics (Princeton, NJ: Princeton Univ. Press)

\bibitem{sofue2009}
Y. Sofue, M. Honma, T. Omodaka, PASJ, 61, 227 (2009)



\bibitem{sofue2013}
Y. Sofue, 
Planets, Stars and Stellar Systems Vol. 5, by Oswalt, Terry D.; Gilmore, Gerard, ISBN 978-94-007-5611-3.
Springer Science+Business Media Dordrecht, p. 985 (2013)


\bibitem{King2015}
C. King III , W. R. Brown , M. J. Geller , S. J. Kenyon
\apj\,,  813, 89 (2015)

\bibitem{Kafle2012}
P. R. Kafle, S. Sharma, G.F. Lewis, \& J. Bland-Hawthorn,  \apj\, 761, 98 (2012)

\bibitem{Deason2013}
A. J. Deason, R.P. van der Marel, P. Guhathakurta, S. T. Sohn, \&
Brown, T. M.  \apj\,, 766, 24 (2013)


\bibitem{Lin2018}
S.-C. Lin, H.-Y. Schive, S.-K. Wong, T. Chiueh, 
\prd, 97, 103523 (2018)

\bibitem{Chen2017}
S.-R. Chen, H.-Y. Schive, T. Chiueh, 
Jeans analysis for dwarf spheroidal galaxies in wave dark matter.
\mnras, {\bf 468}, 1338-1348 (2017)	

\bibitem{chan2018}
J.H.H. Chan,  H.-Y. Schive, T-P. Woo, T. Chiueh, 
\mnras, 478, 2686-2699 (2018)



\bibitem{Xenon1T}
E. {Aprile},  et al., 
\prl, {\bf 121}, 111302 (2018)

\bibitem{Tremaine1979}
S. Tremaine, J.E. Gunn, \prl, 42, 407 (1979)

\bibitem{Oh2011}
S.-H. Oh, W. J. G. de Blok, E. Brinks, F. Walter, R. C. Kennicutt,
\apj, 141, 193 (2011)

\bibitem{Moore1999}
 B. Moore,  S. Ghigna,  F. Governato,  G. Lake, T. Quinn, J. Stadel, P. Tozzi, \apj, 524, L19 (1999)

\bibitem{Boylan-Kolchin2012}
 M. Boylan-Kolchin, J. S. Bullock, M. Kaplinghat, \mnras, 422, 1203 (2012)

\bibitem{Hezaveh2016}
Hezaveh et al. \apj, 823, 37 (2016)

\bibitem{Torrealba2019}
 Torrealba G.,Belokurov V.,Koposov S. E.,Li T. S.,Walker M. G., Sanders J. L., Geringer-Sameth A.,Zucker D. B.,Kuehn K.,Evans N. W.,Dehnen W.,
 \mnras, in press (2019). https://doi.org/10.1093/mnras/stz1624
 
\bibitem{Broadhurst2019}
 T. Broadhurst, I. De Martino, H. N. Luu, G. F. Smoot III, S.-H. H. Tye, arXiv:1902.10488 (2019) 
 
 \bibitem{Viel2005}
 M. Viel, J. Lesgourgues, M. G. Haehnelt, S. Matarrese, A. Riotto,
\prd, 71, 063534 (2005)

\bibitem{Bulbul2014}
 E. Bulbul, M. Markevitch, A. Foster, R. K. Smith, M. Loewenstein, S. W. Randall, \apj, 789, 13 (2014)

\bibitem{Rocha2013}
 M. Rocha, A. H. G. Peter, J. S. Bullock, et al., \mnras, 430, 81 (2013)


\bibitem{Maccio2012}
 A. V. Macci\'o, S. Paduroiu, D. Anderhalden, A. Schneider,B. Moore, \mnras, 424, 1105 (2012)
 
 \bibitem{Hlozek2015}
R. Hlozek, D. Grin, D. J. Marsh, P. G. Ferreira, \prd, 91, 103512 (2015)

\bibitem{Hlozek2017}
R. Hlozek,  D. J. E. Marsh, D. Grin, R. Allison, J. Dunkley, E. Calabrese, \prd, 95, 123511 (2017)

\bibitem{Calabrese2016}
E. Calabrese, D. N. Spergel,
Ultra-light dark matter in ultra-faint dwarf galaxies.
\mnras, {\bf 460}, 4397 (2016)

\bibitem{Marsh2015}
 D. J. E. Marsh,  A.-R. Pop, \mnras, 451, 2479 (2015)

\bibitem{Frye2001}
B. Frye, T. Broadhurst, N. Benitez
\apj, 568,  2, (2001)
 
\bibitem{Pettini2001}
M. Pettini, et al., \apj,554, 981-1000 (2001)
 
\bibitem{Baur2016}
 J. Baur,  N. Palanque-Delabrouille, C. Y\'eche, C. Magneville, M. Viel, \jcap, 2016, 012 (2016)
 
 \bibitem{Irsic2017a}
 Ir\v{s}i\v{c} et al. \prl, 119, 031302 (2017)

\bibitem{Irsic2017b}
 Ir\v{s}i\v{c} et al. \mnras, 466, 4332 (2017)

\bibitem{Nori2019}
M. Nori, R. Murgia, V. Ir\v{s}i\v{c}, M. Baldi, M. Viel
\mnras, 482, 3, 3227-3243 (2019)


\bibitem{Armengaud2017}
E. Armengaud, N. Palanque-Delabrouille, C. Y\'eche, D. J. E. Marsh, J. Baur, \mnras, 471, 4606-4614 (2017)

\bibitem{Garzilli2017}
A. Garzilli, A. Boyarsky, O. Ruchayskiy, \plb, 773, 258-264 (2017)

\bibitem{Schive2019}
Ka-Hou Leong, Hsi-Yu Schive, Ui-Han Zhang, T. Chiueh,
\mnras, 484, 4273-4286 (2019)

\bibitem{Torrealba2016}
G. Torrealba, S. E. Koposov, V. Belokurov, M. Irwin, 
\mnras, {\bf 459}, 2370-2378 (2016) 










 















 
 

























\bibitem{Molinari2011}
S. Molinari et al. \apjl, 735, L33 (2011)

\bibitem{Kruijssen2015}
J. M. D. Kruijssen, J. E. Dale, S. N. Longmore 
\mnras, { 447}, 2, 1059-1079 (2015).

\bibitem{Henshaw2016}
J. D. Henshaw, S. N. Longmore, J. M. D. Kruijssen,  
\mnras Letters, { 463}, 1, L122-L126 (2016)

\bibitem{Khmeinitsky:2013lxt} 
A. Khmelnitsky and V. Rubakov, JCAP, { 2} , 019 (2014).

\bibitem{idm2017}
I. De Martino, T. Broadhurst, S.-H. H. Tye, T. Chiueh, Hsi-Yu Schive, R. Lazkoz,
\prl, 119, 221103 (2017)

\bibitem{idm2018}
I. De Martino, T. Broadhurst, S.-H. H. Tye, T. Chiueh, Hsi-Yu Schive, R. Lazkoz, 
Galaxies, 6, 10 (2018)







\end{thebibliography}
\end{document}